\documentclass[12pt, preprint]{aastex}
\begin{document}
\title{The Star Formation History of the Blue Compact Dwarf Galaxy 
UGCA~290\footnote{ 
Based on observations made with the NASA/ESA Hubble Space
Telescope, obtained at the Space Telescope Science Institute, which is
operated
by the Association of Universities for Research in Astronomy, Inc., under
NASA contract NAS 5-26555.}} 

\author{Mary M. Crone}
\affil{Skidmore College, Saratoga Springs, NY 12866, USA}
\email{mcrone@skidmore.edu}
\author{Regina E. Schulte-Ladbeck}
\affil{University of Pittsburgh, Pittsburgh, PA 15260, USA}
\email{rsl@phyast.pitt.edu}
\author{Laura Greggio}
\affil{Osservatorio Astronomico di Bologna, Bologna, Italy, and
Universit\"{a}tssternwarte M\"{u}nchen, M\"{u}nchen, FRG}
\email{greggio@usm.uni-muenchen.de}
\author{Ulrich Hopp}
\affil{Universit\"{a}tssternwarte M\"{u}nchen, M\"{u}nchen, FRG} 
\email{hopp@usm.uni-muenchen.de}

\begin{abstract}
We present the star formation history of UGCA~290,
a galaxy with properties intermediate between Blue Compact Dwarfs
and Dwarf Irregulars.  This galaxy is particularly interesting because
its young stellar population is extremely similar to that of 
the well-studied type iE 
Blue Compact Dwarf VII~Zw~403, despite its  
different spatial morphology and
old stellar content.  Our Hubble Space Telescope/Wide Field Planetary Camera 2
single-star photometry for UGCA~290 
extends over nine magnitudes in $I$, and allows a 
detailed study of its star formation history. 
Using synthetic color-magnitude diagrams, we show that 
the recent ``burst" which gives this galaxy its BCD status is a moderate 
enhancement
in star formation which lasted for approximately 20 Myr, at a rate 
about ten times above 
its previous rate. 
The star formation history for most of the previous billion years 
is consistent with a constant rate, 
although 
enhancements as large as the current one 
are possible  
at times ealier than
400 Myr ago.  We estimate that the total mass converted into stars in UGCA~290 
more than one billion years ago is about three times the 
astrated mass since that time. 
The initial mass function is consistent with a Salpeter slope, 
and the stellar
metallicity is bracketed by Z$_\sun$/50 and Z$_\sun$/5, with evidence for metallicity 
evolution. 
Similar results for the star formation history over the past 600 Myr 
apply to VII~Zw~403. 
Our main result is that despite the
traditional picture of BCDs, the current bursts in these two galaxies 
are neither remarkably intense nor short-lived, 
and that most of their star formation occured more than a billion years
ago. 
\end{abstract}


\section{INTRODUCTION}

The discovery of two extremely low-metallicity blue dwarf galaxies by Searle \& 
Sargent (1972) launched efforts to determine whether such 
galaxies are young 
``in the sense that most of their star formation has occurred in recent times"
or alternatively, whether ``star formation in them occurs in intense 
bursts which are
spearated by long quiescent periods."   Their primary argument was that the 
current level of star formation
would quickly overproduce heavy elements, at least according to a Salpeter
initial mass function.  
The category of Blue Compact Dwarf (BCD) has since grown to encompass 
galaxies with
a wide range of morphologies, including galaxies identified through  
spectral lines, as well as color and compactness (Thuan \& Martin 1981).  The
categories of HII galaxies and amorphous galaxies, 
while reflecting different selection
criteria, apparently have very similar intrinsic properties to BCDs 
(Marlowe, Meurer, \& Heckman 1999). 

Much of the interest in BCDs springs from the first possibility:  that BCDs are
young.  As such, their existence would support the 
delayed formation of dwarfs scenario, in which ionization of the intergalactic
medium inhibits star formation in systems with shallow gravitational wells
(e.g. Babul \& Rees 1992). 
They would also provide nearby, observationally convenient examples 
of ``primordial" 
galaxy formation.
However, deep CCD imaging in the 1980s 
ruled out the possibility that the majority of 
BCDs began forming stars within the last billion years, 
revealing red elliptical background sheets of presumably older stars 
(e.g. Loose \& Thuan 1986; Kunth, Maurogordato, \& Vigroux 1988). 
Indeed, several of these sheets have now been resolved into red giant stars, 
implying star formation at
least 1-2 Gyr ago (e.g. Schulte-Ladbeck, Crone, \& Hopp 1998; 
Lynds et al. 1998; 
Schulte-Ladbeck et al. 2000; Drozdovsky et al. 2001). 
 Some authors still claim  
that the most extreme BCDs ---  
those with especially low metallicities --- 
formed very recently, even as recently as
100 Myr ago (Izotov \& Thuan 1999, Izotov et al. 2000).  But this 
possibility is dwindling as well;  red background sheets continue to be
discovered (e.g. Pox~186; Doublier et al. 2000), and 
single-star photometry of the 
most low-metallicity BCD on record suggests stars at least as old as 500 Myr
(I~Zw~18; Aloisi, Tosi \& Greggio 1999; \"{O}stlin 2000). 

The alternative possibility of intense repeated bursts is problematic also. 
This scenario requires the existence of a large population of quiescent 
counterparts to BCDs, but no such population has been clearly identified
despite considerable effort 
(Kunth \& \"{O}stlin 2000).  In fact, 
several additional lines of evidence point to the possibility that 
most BCDs are not
really bursting 
in the traditional sense of very short flashes separated by very long
quiescent states. 
Those with resolved stars do not show evidence of long gaps in 
their recent star formation history  
(although a short gap
is possible in I~Zw~18;  Aloisi et al. 1999),  
nor do they exhibit very large star formation rates, only in the range 
$10^{-3}$ to $10^{-2}$ M$_\sun$yr$^{-1}$ (Schulte-Ladbeck et al. 2001).  
Large-scale surveys support the conclusion that star formation rates for BCDs
are typically less than 0.3 M$_\sun$yr$^{-1}$ (Popescu, Hopp, \& Rosa 1999).  
Thus, the nature of the BCD designation, including the details of the current burst
and the relationship to other gas-rich dwarfs with less intense levels of 
star formation, is still a puzzle. 

To address these questions, we obtained Wide Field Planetary Camera
images of UGCA~290, the most nearby BCD without a large red background sheet.
Up to this point, all BCDs resolved as deeply as the red giant branch were
of the common subtype iE, in which the star forming region is embedded in 
a large elliptical background of older, red stars.  Ground-based images
of UGCA~290, on the other hand, show that its star formation extends over
a large fraction of the visible galaxy, 
in the form of two large lobes. 
Single-star
photometry of a BCD with this different kind of morphology provided us the 
opportunity to 
gain insight into the general nature of the BCD designation. 

In a recent Letter, we presented our $V, I$ color-magnitude diagram from
these data (Crone et al. 2000). 
We found a red giant branch, corresponding to a 
spatially compact population of stars older than $1-2$ billion years,  
within which the current star formation is embedded. 
From the magnitude of the tip of the red giant branch we found a distance
of 6.7 Mpc, and distance-dependent parameters M$_B = -13.4$ and 
L$\alpha = 1.2\times10^{39}$ erg s$^{-1}$.  Overall, we found that 
the properties of UGCA~290
are intermediate between those of BCDs and Dwarf Irregulars. 
We also noted that its bright stellar content is extremely 
similar to that of the well-studied iE BCD VII~Zw~403, despite its 
different spatial morphology and old stellar content.

Here we present our full HST photometry of UGCA~290 in the filters
F336W, F555W, F814W, and F656N.  We then investigate its nature using 
synthetic color-magnitude diagrams, addressing the meaning of the BCD designation 
through the details of the current burst and its earlier star formation history. 
Our main conclusion is that neither of the two scenarios proposed by Searle \&
Sargent fits the nearby BCDs UGCA~290 and VII~Zw~403. 

\section{OBSERVATIONS AND REDUCTION}
We observed UGCA~290 in August of 1999 as part of GO program
8122.  
We obtained data in four filters:  F814W, 
F555W, F336W, and F656N.  For F814W and F555W,
which approximate $I$ and $V$, we took six 1300 s exposures, 
at three dither positions, 
for a total of 7800 s in each
filter.  For F336W and F656N we obtained two 1300 s exposures at one position, 
for a total of 2600 s  in each
filter.  

We combined exposures at the same pointings using CRREJ and then combined the 
dithered images using DRIZZLE onto a $1600\times1600$ pixel grid. 
The drizzling procedure allowed us to recognize and   
remove cosmic rays left by CRREJ (which have a characteristic basketweave shape 
in the drizzled image because they do not appear in all three dither positions)  
and hot pixels missed by WARMPIX (which appear as three diagonally
spaced hotspots.)  
Figure~1 shows the color image of the PC, along with a ground-based image we
obtained at the Calar-Alto 1.23 m telescope.  We captured most of the galaxy 
within the WFPC2 field,
but missed part of the lower surface brightness region to the northeast. 
Note that the galaxy is quite transparent --- see the
two red background galaxies --- and free of bright HII regions. 
The two lobes of star formation which give the galaxy a bimodel appearance
are the large regions in the upper left and lower right. 

After masking out a few obvious background galaxies,
we conducted single-star photometry
with DAOPHOT.  We took the zero points from the May 1997 SYNPHOT tables and 
determined our point spread function (PSF) from relatively isolated 
stars in our images.  With no drizzling, the full width at 
half maximum of the 
PSF is 0.07 arcseconds
for the PC chip and 0.13 arseconds for the WF chips, in each filter.
The drizzling procedure for the F555W and F814W filters produces better
defined but marginally larger PSFs, with a full width at 
half maximum of 0.08 arcseconds for the PC 
chip and
0.15 arcseconds for the WF chips. 
Figure 2 shows the DAOPHOT errors from the PSF fitting procedure. 
 Note that we do not 
see as deeply in the F336W filter, partly because of our shorter exposure
time and lack of dithering in this filter.   

We estimated the completeness of our photometry by adding artificial stars to our images
using ADDSTAR, and checking to see what percentages of them we could recover.  In order
to maintain the same level of crowding as in the original, we created simulated images
by adding 5\% of the observed number of resolved stars,
consistent with their observed luminosity function.  We created 200 such images for
each filter in two different regions: in the PC chip, 
to model the more crowded inner part 
of the galaxy; and in the quarter of the WF2 chip closest to UGGA~290, 
to model the background sheet. 
We considered a star to be recovered if the
difference between its input magnitude and recovered magnitude 
was less than 0.7 mag.  Figure 2 includes the results of these tests. 
In most cases, the completeness for the more sparsely populated 
WF2 chip is better than for the PC
chip, despite its inferior resolution; only for bright stars in the 
F336W filter is this trend significantly reversed. 
Completeness tests also provide an estimate of error: 
the difference between the added magnitudes and recovered magnitudes
of the false stars (Figure 2, bottom row).
Note the systematic brightening at faint magnitudes, presumably because of
blending with nearby stars.  This effect cannot be modeled using
simply the DAOPHOT errors, which provide an rms value only. 
DAOPHOT errors are also less likely to take into account correctly
stars which overlap
concentrically.  For these reasons, we use the false star errors
rather than the DAOPHOT errors in the analysis which follows. 

We corrected for the small foreground extinction in this direction according to 
Schlegel et al. (1999), assuming an $R_{\rm V} = 3.1$ extinction curve: 
$A_{\rm U} = 0.076$, $A_{\rm V} = 0.046$, $A_{\rm R} = 0.037$, 
and $A_{\rm I}=0.027$. 
We have not attempted
to correct for extinction within UGCA~290, but 
its transparency outside major regions of star formation
suggests that it is low, at least for stars in the background sheet. 
Finally, we transformed our F555W and F814W magnitudes into 
ground-based $V$ and $I$
following Holtzman
et al. (1995).  We chose not to transform the F336W band into $U$, 
because of the larger uncertainties
in this transformation (Holtzman et al. 1995).  
Finally, we cross-identified stars among different filters requiring  
spatial coincidence 
within the full width at half maximum of the PSF. 

\section{STELLAR CONTENT} 
Our deepest images come from the long, dithered exposures  in $V$ and $I$. 
Figure 3 shows our $V, I$ color-magnitude
diagram (CMD), along with that of the well-studied iE BCD VII~Zw~403.  
(Our reduction of the VII~Zw~403 data
is described in Schulte-Ladbeck et al. 1998).  
The CMD for each galaxy 
includes the entire region of active star formation and most of
the red background sheet, and 
therefore approximately 
reflects the star formation history of the entire galaxy. 
To help interpret the nature of the hottest stars in the diagram, whose
$V-I$ colors are less sensitive to temperature, 
we incorporate information
from the F336W filter as well.  Stars with F336W-F555W color less than $-1.0$ 
are colored blue, those between $-1.0$ and 0.0 green, 
those between 0.0 and 1.0 yellow,
and those greater than 1.0 red.   
Stars in black were not detected in F336W.  
The absolute
luminosity scale is set by the magnitude of the tip of the red giant
branch (TRGB), as described in Crone et al. 2000.  

Several 
features are clear in both CMDs.  The abundance of very blue stars 
($V-I<0$) suggests
a strong main sequence up to $M_I = -6$, while the 
slight redward turn
of the blue plume at very bright magnitudes suggests post-main sequence
stars older than a few Myr.  Unfortunately, there is no clear separation
in the lower blue plume between main sequence stars and blue He-burning
stars. 
There is a mix of F336W-F555W colors 
within the blue plume.  For example, among the very brightest 
stars in UGCA~290
are ones color-coded both yellow and blue.  This effect is not simply
due to photometric errors, which 
are less than 0.2 mag in F336W-F555W for these stars; it probably reflects
the presence
of stars in different evolutionary phases coexisting on the $V, I$ CMD. 
It is also possible that some stars with anomalous colors are actually
binary systems of bright, differently colored stars. 
The blue plume population for VII~Zw~403 contains a slightly larger population
of very hot stars (color-coded blue) than that for UGCA~290. 
A well-defined red plume at $1.0<V-I<2.0$ peaks at about the same
magnitude as the blue plume, and is smoothly populated down to the tip
of the asymptotic giant branch at M$_I \sim -5$.  The red giant branch 
occupies the region redward of $V-I \sim 0.5$ and faintward of M$_I \sim -4$.  
In the case of VII~Zw~403, the red giant branch
is very heavily populated, and there are many asymptotic giant branch (AGB) stars 
between M$_I=-4$ and $-5$, magnitudes expected for ages between about 600 Myr and
1200 Myr.  This large older population occupies an elliptical sheet 
several kpc across, surrounding the compact ($< 1$ kpc) star forming region 
(Schulte-Ladbeck et al. 1999).  This is 
typical of the type iE BCD morphology.  In contrast, a 
relatively small number of these older stars appear in UGCA~290, and 
their spatial distribution is much smaller relative to the star forming
region (Crone et al. 2000). 
Despite these differences in older populations, however, the younger
populations are clearly similar.  

To quantify this similarity, we performed a two-dimensional 
Kolmogorov-Smirnov (K-S) test 
using the subroutines in Press et al. (1992), with significance
levels determined using a bootstrap normalization for our particular
data (see, for example, the procedure in Fasano \& Franceschini 1987). 
A fair comparison of the two CMDs using this test 
requires that errors and completeness issues
do not affect the two distributions differently; 
we limited the data to the
region M$_{\rm I}<-4$, M$_{\rm V}<-3.7$, for which completeness is better than 
$95\%$ and photometric errors are smaller than 0.09 in each filter. 
This region includes 424 stars for UGCA~290 and 408 stars for VII~Zw~403.
(Note that the K-S test is designed to find differences in distribution rather
than total number of points.) 
We find that the stars in this region, which stretch over nearly five magnitudes,  
are not distinguishable using the K-S test. 
If we extend the comparison region of the CMD to include 
even a small bit of the older
stars in the red giant branch or asymptotic giant branch, the K-S test easily
distinguishes the distributions.  For example, shifting the cutoff down to 
M$_{\rm I}<-4$, M$_{\rm V}<-3.0$ results in distributions that are different at
the 99.9\% level.   

Another simple quantitative measure of the two stellar populations  is
their luminosity function.  Figure 3 includes the completeness-corrected  
$I$-band luminosity functions of the blue and red plumes, split in
color at $V-I = 0.7$. 
We fit each histogram 
to the function
$ {\rm log}N = {\rm log}N_{\rm o} + \alpha M_{\rm I}$, again limiting 
ourselves to magnitudes where
the photometry is better than 95\% complete 
($M_{\rm I} < -4$ for the blue plume 
and $M_{\rm I} < -5.5$ for the red plume).  
In all cases, this function provides a good
fit, with a reduced chi-squared of less than 1.5.  
As expected from the
K-S test results, the
slopes for the two galaxies are consistent with each other. 
For the blue plume, the slopes $\alpha$ are $0.44\pm0.03$ for UGCA~290
and $0.49\pm0.03$ for VII~Zw~403.  For the red plume,
they are $0.22\pm0.05$ and $0.26\pm0.06$, respectively. 
The histograms also highlight the slightly greater population of bright
stars in UGCA~290 brighter than M$_{\rm I}=-6$, 
and the significantly greater populations of red giants and asymptotic giant
branch stars in VII~Zw~403. 
The total number of stars in the faintest magnitude bins are extremely 
sensitive to completeness estimates and are therefore difficult to compare.
 
Figure 4 presents our full results for F336W, potentially valuable not only
for interpreting the hottest stars but also for estimating extinction. 
Before discussing this figure, we caution that for several reasons 
the F336W data are more difficult to interpret than the data in F555W 
and F814W.  As noted above, the transformation between F336W and $U$ is
rather uncertain, depending on stellar surface gravity and, 
for redder stars,
the red leak in the F336W filter.  
Combined with our shorter exposure
time and lack of dithering, these limitations prompt us to focus our
quantitative study on $V$ and $I$, and view the F336W information as
supplemental.   

Still, the F336W filter can provide some useful 
information.  
In Figure 4, color-coding indicates M$_I$ magnitude, and is helpful for
identifying stars with their position on 
the $V-I$ color-magnitude diagram. 
 Stars with M$_I < -8$
are represented in blue, $-8<M_{\rm I}<-6$ in green, $-6<M_{\rm I}<-4$ in
yellow, and $M_{\rm I}>-4$ in red.   
The bright end of the blue plume in F336W, F555W is better defined for 
VII~Zw~403; note how the bright (e.g. green-colored) 
stars are more concentrated
at blue F336W-F555W color.  This corresponds to 
the higher proportion of very hot,
bright stars in VII~ZW~403 visible 
in Figure 3.  The K-S test applied to the region above the dashed line 
in Figure 4 distinguishes 
the bright populations to better than 99\% certainty. 

In the color-color diagrams, 
the main features are the cluster of mostly fainter (red-colored) stars
at  F336W-F555W $\sim -1.5$,
$V-I \sim -0.2$  and the tail of mostly brighter stars extending downward. 
Note that the cluster of faint stars extends to redder F336W-F555W colors
in the case of VII~Zw~403;  this is because the F336W data are deeper for VII~Zw~403,
allowing us to see the fainter, redder stars at M$_{\rm V}\sim-3.5$, 
F336W-F555W$\sim-1.3$. 
The arrows show a reddening vector for $E(B-V) = 0.5$, according to 
the extinction curve of Cardelli, Clayton, \& Mathis (1989).  
Lynds et al. (1998), compare their color-color 
diagram of VII~Zw~403 (based on these same data) with a 
theoretical isochrone for 
metallicity Z=0.008 and age 4 Myr, transformed into the F336W, F555W system
according to Holtzman et al. (1995). They conclude that there is very little 
internal reddening in this galaxy.  (Note that we do find a stream of about 
a dozen stars
coming from the main cluster of VII~Zw~403 in the direction of the reddening
vector, which probably does indicate substantial reddening for these few
stars.) 
 Given the similarities of these two
color-color diagrams, the same result would hold for UGCA~290. 
In some cases it is useful to determine the differential reddening of each star and
then correct the CMDs for this effect.  We prefer
not to do this.  For one thing, these stars are in different, 
unknown evolutionary phases.
For example, the colors of the three stars at $(V-I)_o \sim 1.5$  could be 
either negligibly reddened red plume stars or heavily reddened main 
sequence stars.   
And again, 
the transformation to the F336W 
filter is rather uncertain.  In any case, there is very little
evidence for reddening. Along the downward tail, for example, there is only
about 0.3 mag of scatter. 
Photometric errors alone reach about 0.1 in
the F336W filter for these stars, and as high as 0.5 mag
for the fainter stars in the main cluster. 

For the purposes of our theoretical modelling, we will focus on our superior
images in $V$ and $I$, with the following input from the F336W filter:  little
evidence of systematic reddening, a scatter indicating differential reddening 
of less than E(B-V)$\sim 0.3$, with a few outliers; 
and a mix of stars with different temperatures
along the blue plume.  

\section{COMPARISON TO SYNTHETIC COLOR-MAGNITUDE DIAGRAMS} 
\subsection{Procedure}

To interpret the star formation history in more detail, we employ synthetic
color-magnitude diagrams which include theoretical stellar evolutionary
tracks and atmospheres, convolved with errors and completeness 
fractions to mimic
our data.  Specifically, we use the Bologna code (Greggio et al. 1998) with 
the Padova tracks (Fagotto et al. 1994) and the atmospheres of Bessell,
Castelli, \& Pletz (1998).    

Estimates of errors and completeness are provided by false star tests,
as described in Section 2.  There are some sources of error, however, 
that are not included in these tests. 
One is differential reddening, which is
likely to cause some scatter in the CMD towards redder colors and fainter
magnitudes.  We do not expect this to be a large effect for this 
particular galaxy, because of its high galactic latitude, its intrinsic
transparency (Figure~1), and because there is little evidence for it
in our color-color diagrams (Figure~4).  
Another is 
the clustering properties of the stars, including 
the importance of compact young clusters and binaries. 
Although we perform completeness tests for both a more crowded ``star formation"
region and a less crowded ``background sheet" region, we do not 
lay down stars precisely according
to their actual clustering properties.  In fact, no matter how careful 
the analysis, one cannot perfectly determine the clustering properties of
the stars directly from the image.

Recently some authors have presented methods to automate the process
of finding the best-fit synthetic CMDs (e.g.  
Harris \& Zaritsky 2001.)
These methods are suited to situations where theoretical predictions
and observational errors are very well quantified, such as 
high-resolution images of nearby galaxies, and well-understood features on the
CMD.  Because our best data are for very young and metal-poor stars
with relatively poorly constrained positions on the
CMD and uncertainties which are difficult to quantify, 
we prefer to use a more ``hands-on" approach. 
Our basic procedure is to consider a series of boxes in color and magnitude
which include stars of
progressively older ages, as described in detail below.  
We used the same kind of procedure in Schulte-Ladbeck
et al. (2001).  Similarly, we do not expect to find very a precise solution for
initial mass function (IMF) and metallicity, but
instead expect to find a fairly wide range of these parameters
consistent with the data.
 
In order to examine the 
likely range of metallicities for this galaxy,  we consider a low value of 
Z=0.0004 (Z$_\sun$/50,
the lowest measured HII region abundance), and a higher value of 
Z=0.004 (Z$_\sun$/5). 
There are no published nebular metallicities for this galaxy,
so this choice of upper limit is guided by
the appearance of the CMD.  
As illustrated in Figure 5a, the stellar populations are likely to 
be bracketed by these two values, with the bright young stars closer 
to Z=0.004 and the older red giant branch stars closer to Z=0.0004. 
Indeed, in Crone et al. (2000) we estimated from the red giant branch
a low stellar metallicity [Fe/H]$=-2.0 \pm 0.1$, which in turn suggests 
a low nebular metallicity $12+{\rm log}$(O/H)$ \sim 7.6$, or Z=0.001. 
We can also use the fact that the young stars have very
similar colors 
to those in VII~Zw~403, which has a measured nebular metallicity (O/H) between 
Z$_\sun$/20 and Z$_\sun$/10 
(Martin 1997; Izotov, Thuan, \& Lipovetsky 1997; Izotov \& Thuan 1999).
Finally, the simple fact that UGCA~290 is as faint as $M_B=-13.4$ 
also suggests $12+{\rm log}$(O/H)$ \sim 7.6$ (Skillman, Kennicutt,
\& Hodge 1989). 
For each metallicity we consider a power law IMF, 
$\xi \propto M^{-\alpha}$ with the slope $\alpha$ in the
range $1.35$ to $3.35$, 
the standard Salpeter IMF (Salpeter 1955) being $2.35$.  The star formation
rates we quote in this section assume a constant slope from 0.1 M$_\sun$ to
100 M$_\sun$.

We model the star formation history using boxes such as those
illustrated in Figure 5b.  The basic idea is to consider a series of 
regions on the CMD 
which include stars of
progressively older ages.  The choice of box location and size is guided by identifying 
regions which have both  
well-determined ages and large enough populations to 
provide a precise estimate of the SFR. To identify such regions, we performed a 
series of simulations showing the appearance of an aging coeval burst. 
We found, for example, that for the Z=0.004 model the stars in Box B
are between about 10 and 15 Myr old, so that the total number of these
stars can be used to determine the star formation rate over this time period. 
The stars produced along with those in Box B must then be taken into account
when considering the next box. 
We used the boxes in Figure 5b 
to determine the star formation rates for the Z=0.004 models
in the periods $0-10$ Myr ago (Box A),
$10-15$ Myr ago (Box B), $15-20$ Myr ago (Box C), $20-50$ Myr ago (Box D), 
$50-1000$ Myr ago (Box E), and more than 1 Gyr ago (Box F).  Note that
although we use Box A, which includes the tip of the main sequence,   
to constrain the youngest population,  this box may also contain stars as old as 
100 Myr; therefore, during the modeling procedure we use it after Box E.  
Note also that 
the time resolution goes down dramatically for Boxes E and F, because 
populations 
with a wide range of ages (and which are not well-constrained elsewhere on the
CMD) cohabit these regions of the CMD.  
Finally, random fluctuations produce a wide range of star formation
rates for boxes with few stars. Not all of these rates are 
consistent with other parts of the CMD --- for example, some random
realizations based on Box B overproduced the lower part of the
main sequence at M$_{\rm I}\sim -4$.  
Only those rates which produced acceptable results overall contributed to 
our determination of the star formation rates and the  
uncertainties in these rates.  We do not use the very faintest magnitudes,
those with completeness estimates of less than 50\%, to constrain the models. 
 
\subsection{The Star Formation History:  Basic Features} 

Before discussing the star formation history in detail, we describe some
overall features of our results. 
The model which best matches our data has a Salpeter IMF and a
metallicity which evolves from the
lower value $Z_\sun$/50 for the RGB stars to the higher value 
$Z_\sun$/5 for the youngest stars.  
Figure 6 illustrates the SFH over the past Gyr for this model.  
The burst peaks $10-15$ Myr ago 
at $0.042 \pm 0.010 {\rm ~M}_\sun {\rm yr}^{-1}$, and then decreases 
to about one fourth of this rate. 
Before the burst, the rate is about one tenth of this. 
Each bin in Figure 10 represents the average SFR determined using one of the
boxes in Figure 5.  The bins for relatively recent times represent
boxes with good time resolution but fewer stars, resulting in 
narrower bins with larger errors bars.  The bins for 
later times represent boxes with many stars of a wide range
of ages, cohabiting the same part of the CMD.  Each bin  
represents a kind of average for the SFR within the time period under
consideration;  there can be unresolved fluctuations in the SFR within these
periods. 
We do not include results for times earlier than 1 Gyr, because we
lose nearly all time resolution at this point.  Section 4.5 below
describes the constraints we can put on the SFR prior to this time. 

The differences between our best model and the data are not
surprising given the 
known uncertainties. 
Figure 7 shows one of the random realizations of our best model. 
This figure excludes stars with $V-I<2.0$ in an attempt to exclude
thermally-pulsing AGB stars from our analysis; 
the simulator does not place these stars on the CMD because the
colors and lifetimes in this phase are not theoretically well-determined.
The data scatter to redder colors than the theoretical model, 
an effect expected from differential 
reddening.  This trend is especially obvious for 
the very brightest supergiants, whose theoretical colors 
are also affected by
uncertainties in the theoretical temperatures for massive post-main sequence
stars (Renzini et al. 1992). 
Second, there are slightly more very faint stars in the data.  These low
luminosity bins are very sensitive to incompleteness and could well represent
an inadequacy in our completeness estimate.  Alternatively, it could reflect a 
slightly steeper luminosity function, or an enhancement in star formation 
several 100 Myr ago, possibilities which we discuss below.  Third,
there is a discontinuity in color in the blue plume at M$_I \sim -6$. 
There are two effects which likely
come into play to ``smooth out" this feature.  Differential reddening
(and any additional blending we have missed) would tend to smooth out such sharp
features.  
Indeed, both synthetic plumes are sharper
than the data.  Also, the relative time spent in the blue and red parts of
the blue loops is quite   
model-dependent (see Renzini et al. 1992).  Figure 8 illustrates the
dependence on metallicity by showing a 
model using 
low-Z tracks only.  In contrast to the high-Z model, this one {\it overproduces} 
blue-loop stars in the blue
plume around M$_I \sim -6$.  A  metallicity slightly lower than Z=0.004 
might therefore
explain the lack of stars here in our best model.  There are 
clearly other difficulties with the low-Z model. 
Most strikingly, it cannot reproduce
the brightest stars because of their very short lifetimes;
even with an IMF as top-heavy as 1.35 
we were still unable to reproduce the very bright
stars without vastly overproducing the fainter stars.  

Our best model has a Salpeter IMF.
A slope as steep as 3.00 clearly overproduces 
faint, blue main sequence stars relative to the brightest supergiants 
(Figure 9). 
Similarly, a slope as flat as 1.35 clearly fails to produce enough
main sequence stars relative to the brightest supergiants (Figure 10). 

To summarize, the basic features of the CMD support our guess that
the metallicity is bracketed by Z=0.004 and Z=0.0004, with the young stars
closer to Z=0.004. 
The IMF slope is close to Salpeter, and is certainly 
bracketed by 1.35 and 3.00.
In the detailed description of the SFH which follows, we will describe the 
robustness of our results with the respect to this range
of parameters. 

\subsection{The Recent Starburst} 
The recent burst peaked $10-15$ Myr ago 
at $0.042 \pm .010~{\rm M}_\sun {\rm yr}^{-1}$, and then decreased 
to about one fourth of this rate. 
Before the burst, the rate was about one tenth of this. 
Each of these features in the SFH can be understood in terms of 
general features 
in the CMD.
The very recent level must produce 
a main sequence as bright as $M_{\rm I}\sim-6$  while not overproducing very bright, 
blue supergiants relative
to red supergiants. 
The intensity and duration of the enhancement, as well as those of the pre-burst
phase, must produce a smoothly populated
red plume brightward of $M_{\rm I} \sim -6$.  If the burst had started earlier, for 
example, the red plume below $M_{\rm I}\sim -8$ would be
overpopulated (Figure 11).  If, on the other hand, the SFR had been constant
at the rate 50 Myr ago, with no burst at all, 
there would not be nearly enough main sequence stars
or bright supergiants (Figure 12).
 
Models with different IMFs result in slightly different burst parameters. 
A steeper IMF, which has a relatively greater number of 
low mass stars, is skewed toward relatively higher rates for
recent times, because
these rates are constrained by more massive stars on the CMD.
For $\alpha=3.00$, the burst rises from $0.3 \pm 0.1~{\rm M}_\sun {\rm yr}^{-1}$ 
in the interval
$15-20$ Myr ago, peaks at $0.6 \pm 0.2~{\rm M}_\sun{\rm yr}^{-1}$ 
from $10-15$ Myr ago, then 
decreases to one third of the peak value;  before the burst, the rate is
only about 1/20 the burst rate. 
Therefore, the burst is 
effectively shorter and more intense. 
This should be taken as an extreme limit to the burst intensity.
For the flatter slope $\alpha=1.35$, the effect is the opposite.  The 
burst remains constant over the $10-20$ Myr period at $0.01~{\rm M}_\sun{\rm yr}^{-1}$,
then decreases 
to a current rate of one eighth this value;  before the burst
the rate is 
just five times lower than the burst rate.   This should be seen as a lower
limit to the burst intensity.  In both cases, extending the burst as far back
as 50 Myr results in an unacceptable buildup of stars in the red plume. 

The low-Z model does such a poor job producing the very brightest 
stars that it does not provide much meaningful input on burst
parameters.  It does provide one very general result, however:  as with
the other models, if the
burst continues as long as 50 Myr, there is an 
unacceptable buildup of stars in the red plume. 

\subsection{The Intermediate Star Formation History} 

Over the long period 
from 1 Gyr ago all the way up to 50 Myr ago, 
the data are consistent with a constant star formation rate, 
at a rate about one half
that of the pre-burst phase (one third for the steep IMF and the same rate 
for the flat IMF).  However,
moderate fluctuations can
easily ``hide" in the data. 
We performed simulations in which we added a burst similar to the
recent one at various times in the past, to see if we could detect them. 
Specifically, we added a 20 Myr burst at   
0.04 M$_\sun$ yr$^{-1}$ which begin at 100 Myr, 200 Myr, 300 Myr, 400 Myr, 500 Myr, 
and 1 Gyr.
While the two most recent times clearly conflict with the data, a burst as recent
as 300 Myr ago effects primarily the lowest magnitude bins, 
and at 400 Myr ago,
the differences could be interepreted as inadequate completeness estimates
(Figure 13).  
A burst at 500 Myr adds only 100 stars to the synthetic CMD, 
and one at 1 Gyr adds only 30.  
Remember that even our ``best model" slightly overproduces faint stars, 
an effect which 
we attributed to incompleteness in our discussion above.  If, instead, we trust  
our completeness estimate totally, this difference could mean a reduced SFR
somewhere in the period 200-500 Myr ago.  We could, for example, decrease the
SFR by a factor of one half for the period 450-500 Myr ago.  

Likewise, we cannot rule out periods of total quiescense.
Although significant gaps in star formation 
are ruled out for more recent times --- we find that gaps 
as large as 10 Myr within the past 50 Myr 
clearly disagree with the data --- large gaps may have occurred at earlier times.
Figure 14 illustrates an extreme example.  In this model there is no star
formation in the period $500-1000$ Myr ago, and yet the model CMD matches the
data nearly as well as our best model.  There is only a slight 
overproduction of 
faint blue Helium-burning stars, because we raised the star formation rate
slightly in the period $50-500$ Myr ago to compensate for the loss of 
red plume stars otherwise produced $500-1000$ Myr ago.  A similar type of
compensation allows a 100-Myr gap as recently as the period $100-200$ Myr ago.
These uncertainties highlight the difficulty in determining a precise 
SFH for intermediate times without an unambiguously identified 
blue He-burning branch.

\subsection{Early Star Formation History} 

A small part of the UGCA~290 red background sheet 
extends into the WF2 chip,
and a slightly larger part of the background sheet is missed by the WFPC2
entirely 
(see Figure 1, and Karachentsev \& Makarov 1998).  Therefore, the results 
for this phase
of the SFH require modeling of data in both the PC and WF chips. 
We illustrate our modeling of this older population using the WF data,
because these are less contaminated by young stars. 
Figure 15a 
shows the CMD for the quarter of the WF2 chip nearest the galaxy center, 
inluding 
300 stars.  There is no indication
that the galaxy extends into other parts of the WFPC2 field of view.  
Far from the main body of the galaxy there are only a few stars, in 
a random spatial distribution and with intermediate $V-I$ colors atypical
for the rest of the galaxy. 
We use the farthest quarter of the WF3 chip to provide an estimate of the
contamination in the WF2 (Figure 15b).

We model the WF2 data using the errors and completeness estimates 
specific to this
chip.  The main feature is the RGB, but there are a few other stars, 
beyond those expected from contamination, 
which suggest a low level  
of more recent star formation.  We do not see deeply 
enough down the RGB to
distinguish details of the SFH earlier than 1 Gyr ago.  Instead we
consider four different models with constant star formation  
at a rate fixed 
to produce the right number of
stars at the TRGB (specifically, within Box F of Figure 5).  
We model two different starting times:  an ``old" model,
starting 10 Gyr ago, and a ``young" model, starting only 2 Gyr ago.  
We use both the ``high" metallicity Z=0.004 model and the ``low"
metallicity Z=0.0004 model. 
Figure 16 shows an example for each set of parameters.  In addition to
the stars older than 1 Gyr, we include in each diagram the 
``contamination" stars from Figure 15b and the results of a simulation
with constant
star formation at 0.0001 M$_\sun$ for the past 1 Gyr. 

To aid in comparing the models, we also plot in Figure 16 the distribution
in color at the TRGB,  within the region $0.8<(V-I)_o<2.3$, $-4<M_I<-3$. 
The low-Z models successfully match the blue edge of the TRGB, but
do not extend to red enough colors.  The high-Z models, on the other hand, match
the red edge of the TRGB but do not extend to blue enough colors. 
Thus, the width of the TRGB is larger than one would expect simply
from our photometric errors.  The most likely 
explanation is metallicity evolution.  It is also possible that 
differential reddening contributes to this effect. 
The models clearly rule out the possibility that the bulk of the
stars are both old and have the higher metallicity, although we cannot
rule out old low-Z stars or
young high-Z stars.  Unfortunately, we cannot determine from these
data when UGCA~290 first
began to form stars.  

Regardless of the time when star formation first began, we can address the
question of whether the galaxy is ``young" in the sense that it formed
{\it most} of 
its stars recently --- for example, within the past 1 Gyr. To do this we compare
the total astrated mass before and after 1 Gyr ago, for the entire galaxy. 
This is a fairly straightforward application of our CMD models. 
Using the SFRs cited above, the total astrated mass within the last 1 Gyr
comes to $2.8\times10^6 ~\rm{M}_\sun$ for the PC with a negligibly small 
contribution from
the outer parts of the galaxy.  The total astrated mass earlier than 1 Gyr ago
is in the range $4-10 \times10^6~\rm{M}_\sun$ in the PC and 
$ 1-2 \times10^6 ~\rm{M}_\sun$ in the WF2.
The higher masses correspond to higher metallicity and higher age. 
As already noted, our HST observations miss the northeast edge of the red 
background sheet, a region about half the area of the PC chip.
If we assume that the population on 
this edge of the background sheet is  
the same as that on the southern edge in the WF2,  
we can scale our results
using the flux in our R-band ground based image.  
Using the region on the 
ground based image farthest
from the galaxy to estimate a background correction,
we find that the flux of the part missed by the WFPC2 is 1.6 to 2.0 times the
flux in the region we see in the WF2.  Therefore, to get the total star formation
rate of the galaxy, we add the value for the PC chip to  
$2.6-3.0$ times the value for the WF2 chip.  The total astrated mass 
more than a billion years ago is
thus $7-16 \times10^6 ~\rm{M}_\sun$ , which is two to six times 
that since. 
For the $\alpha=3.00$ models, where rates are skewed higher for
recent times relative to the Salpeter IMF models, 
the astrated mass before a billion years ago 
is still two to three times higher 
than that since.  For $\alpha=1.35$, where rates are skewed lower for
recent times,  the astrated mass before a billion years ago 
is five to twenty times
higher than that since.  If star formation began earlier than 10 Gyr ago, 
the astrated mass is higher than otherwise.  For example,
For $\alpha=2.35$ and star formation beginning 15 Gyr ago, the astrated mass
before a billion years ago is six to nine times higher than that since. 
Therefore, UGCA~290 formed most of its stars more than a billion years ago.
In this sense, it is not ``young." 

Another way to compare quantitatively the early SFH with the more
recent SFH is 
through the birthrate parameter ${\rm SFR}/<{\rm SFR>}_{\rm past}$, where the numerator
represents the average SFR over a recent time period --- for example,
the past billion years ---  and the denomenator represents the average SFR
earlier than this (see Schulte-Ladbeck et al. 2001).  In these terms,
the birthrate parameter for UGCA~290 ranges from about 0.4 to about 
4.0 depending on age, metallicity and IMF.  For a Salpeter IMF and 
the low metallicity Z=0.0004, 
it is 0.5 for a 2-Gyr old galaxy and 2.5 for a 10-Gyr old galaxy.
If we extend the SFH back to 15 Gyr ago, the birthrate parameters are
about the same as for the 10-Gyr case;  
although the primary effect of lengthening the period of star formation is
to reduce the average rate, there is also the competing effect that the total
astrated mass is higher in models where star formation begins earlier.  
For comparison, our modeling
of HST/Near Infrared Camera photometry for the BCD galaxies Mrk~178 
(Schulte-Ladbeck et al. 2000) 
and
I~Zw~36 (Schulte-Ladbeck et al. 2001) both produced 
slightly larger birthrate parameters
on the order of ten.  

\subsection{Summary of Synthetic Modeling}
We consider the following our robust results:  
The recent ``burst" which granted UGCA~290 its BCD status is an enhancement of 
about ten times the previous star formation rate during the past 1 Gyr, 
peaking at a rate of $0.04\pm0.01~{\rm M}_\sun {\rm yr}^{-1}$.
It has lasted for about 20 Myr,
and not longer than 50 Myr.
The current star formation rate is likely to be a few times lower than the
peak of the burst. 
Over the long period from 1 Gyr up to 50 Myr ago,  the SFR is consistent
with a constant rate, but modulations as large as the recent burst can
hide for times earlier than 400 Myr ago, as can quiescent periods 
hundreds of millions of years long.  
The blue side of the tip of the red giant branch agrees with a 
low metallicity Z$=0.0004$, while its red extension suggests 
metallicity evolution.  The number of stars at
the TRGB, scaled to the entire galaxy using the R-band image, indicates that
the total mass astrated before 1 Gyr ago is several times 
the total astrated mass since then. 

\section{SPATIAL DISTRIBUTION OF THE RECENT ENHANCEMENT} 

There are several questions about the history of UGCA~290 which can be addressed
by the spatial distribution of resolved stars, 
including whether there is any evidence of
propagating star formation, and 
whether the galaxy represents an interacting
pair or recent merger. 
Figure 17 illustrates the spatial distribution of stars of progressively
older ages.  From left to right, the stars were selected to have formed 
during 
the burst (the past 20 Myr), the preburst phase (20 to 50 Myr ago), 
the intermediate period (50 Myr to 1 Gyr ago), and the early 
period (more than 1 Gyr ago).  We have attempted to select as many stars 
as possible without
contamination by stars of different ages.  Based on our best model,
this means 
$M_{\rm I}<-7.5$ (very bright supergiants) or $M_{\rm I}<-4, V-I<-0.2$ (main
sequence stars); $-7.5<M_{\rm I}$ and $M_{\rm I}<-6.0$, $V-I > 0.2$; 
$-6.0<M_{\rm I}<-4.5$ and $V-I > 0.2$; and $M_{\rm I}>-4.0$ and $V-I>1.0$.  The last
category may include AGB stars as young as 500 Myr along with the older stars. 
The most obvious trend, not suprisingly, is that the older stars
are more dispersed.  
The structure of the burst is clearer in Figure 18, where we zoom in on 
the star forming region. 
The two large lobes 
are obvious (the northern lobe
looks like a double cluster),   
but there are also more widely distributed stars, which
arc between the two main lobes, forming two bubble-like shapes.  
On the right panel of Figure 18 we overlay the burst stars 
with the diffuse $H_\alpha$ distribution 
from our F656N images.
To produce contours of the diffuse gas, we subtracted point sources 
from the continuum-subtracted image and smoothed to
a scale of 30 pc.  
The large-scale H$\alpha$  
morphology of UGCA~290 is clearly 
different from that in more typical BCDs such as 
VII~Zw~403, which show 
bright knots surrounding young stars (e.g. Crone et al. 2001). 
Instead, the burst stars appear on the edges of two large regions with 
diffuse emission.  There are no bright stars visible within the most intense 
knots of H$\alpha$ emission, despite their relatively low surface brightness.
It is tempting to guess that the young stars have blown away
the gas out of which they formed.  

To address the relationship of the two large lobes of star formation, we
examine their individual CMDs (Figure 19).  The stellar content in these
two regions is similar, the most obvious differences being 
the red AGB stars in the South lobe and the small space in the 
blue plume at $I=23$ in the North lobe. 
The K-S test, applied to the entire 
CMD (including 862 stars in the south lobe and 842 stars in the north lobe),
shows a 10\% probability that they are drawn from the same distribution. 
Thus, there is no compelling evidence that 
star formation is  
propagating from one region to the other, or that they represent 
different objects which are merging or interacting.  
On the contrary, 
the burst seems to be a large-scale phenomemon over the entire 
central region of the galaxy.

\section{DISCUSSION AND CONCLUSIONS} 

Our main result is that the recent ``bursts" in both UGCA~290 and
VII~Zw~403 are not especially 
extreme, and that we see no evidence for long gaps in star formation in
their recent history.  These results are similar to those for
other nearby BCDs resolved with HST.  
The star formation rates we 
determine from these galaxies  are on the order of   
$10^{-3}-10^{-2}~M_\sun~{\rm yr}^{-1}$ (see the review in Schulte-Ladbeck 2001).
Indeed, the only late-type dwarf whose resolved stellar population
indicates a major burst is NGC~1569 (Greggio et al. 1998).  As reviewed
by Tosi (2001), this galaxy has a SFR per unit area of 
$ \sim 4~M_\sun ~{\rm yr}^{-1}{\rm kpc}^{-2}$, which is $\gtrsim 40$ 
times the rate in other nearby late-type dwarfs. 
In these units, UGCA~290 
(with a star forming region about 1~kpc across), 
has $0.01~M_\sun~{\rm yr}^{-1}{\rm kpc}^{-2}$ and VII~Zw~403
(with a star forming region about 0.5~kpc across; Crone et al. 2000) 
has $0.02~M_\sun~{\rm yr}^{-1}{\rm kpc}^{-2}$. 
Likewise, we do not see evidence for major gaps in star formation
in nearby BCDs. 
For example, we find that for a sample of four BCDs observed with HST/NICMOS,
very long
gaps in star formation ($> 1$ Gyr) could not have occurred just prior
to the current burst (Schulte-Ladbeck et al. 2001).    
Therefore, the traditional picture of
intense bursts and long gaps does not seem to apply to nearby BCDs.
This is not entirely surprising. 
The distances to most BCDs are not precisely known, making
it difficult to obtain accurate estimates of their star formation rates.
Moreover, the statistics obtained for relatively distant BCDs are subject to
a Malmquist bias,  producing a higher average
SFR than would a volume-limited sample.  Finally, it now appears
likely that the metallicity in BCDs is significantly affected by
preferential loss of enriched gas and infall of low-metallicity gas
(see Kunth \& \"{O}stlin 2000 for a review).

The star formation history from our CMD modeling is quantitatively
consistent with the few other observations available for UGCA~290. 
From our continuum-subtracted F565N image, its H$\alpha$ flux  
is 
$2.0\pm0.3\times 10^{-13}$ erg s$^{-1}$ cm$^{-2}$, which for an extinction of 0.1~mag 
corresponds to a luminosity of $1.21\times 10^{39}$ erg s$^{-1}$.  
Using the conversion in Hunter \& Gallagher (1986) for a Salpeter IMF from
$0.1 {\rm M}_\sun$ to $100 {\rm M}_\sun$, the star formation rate
is then $0.0085\pm0.002~{\rm M}_\sun{\rm yr}^{-1}$. 
This is 
consistent with the rate $0.011 \pm 0.008~{\rm M}_\sun{\rm yr}^{-1}$ 
we derive here from the resolved stars.  We can also check for consistency between
our astrated mass and the dynamical mass of UGCA~290 from H$\alpha$
and HI spectroscopy.  
For a Salpeter IMF, we derive an astrated mass of 
$9.6-18.4\times 10^6~{\rm M}_\sun$. Considering IMF slopes in the range
1.35 to 3.00 broadens 
this range to $9.6-63.5\times 10^6~{\rm M}_\sun$. If 30\% of
the mass is recycling to the interstellar medium through
stellar winds and supernova, the current stellar mass comes to about
$7-13\times 10^6~{\rm M}_\sun$, and 
possibly as high as $40\times 10^6~{\rm M}_\sun$. 
These estimates are based on a single-slope IMF down to a mass of
0.1~M$_\sun$.  For an IMF which flattens for low masses, as 
observed in the galactic disk, the total
astrated mass decreases.  For example, if the slope between 
$0.1-0.6~{\rm M}_\sun$  
 is $-0.564$ (Gould, Bahcall, \& Flynn 1997), the astrated mass decreases
by factors of 0.9, 0.6, and 0.5 for slopes of 3.00, 2.35, and 1.35, respectively.
The total mass of course includes loose gas as well.  We can estimate
the neutral gas using 
the integrated HI flux of UGCA~290 
S=2.18 Jy km/s (Karachentsev, private communication). 
Using the standard formula 
M$_{\rm H}=2.36\times10^5{\rm d}^2$S, 
where S is the flux in Jy km/s, d is the distance in Mpc, 
and the mass is in solar masses,
we obtain $2.3 \times 10^7 {\rm M}_\sun$.  
This is several times higher than the 
stellar mass for a Salpeter slope. 
We can also make a very rough estimate of the ionized gas mass 
from the H$\alpha$ flux.
Following Clayton (1987), and assuming a gas volume of 1 kpc$^3$ and a filling
factor of 0.1, the mass in ionized hydrogen is about $3\times 10^4{\rm M}_\sun$, 
which is negligible compared  to the other sources of mass.
A lower limit to the total dynamical mass can be derived from 
the rotation curve. 
Using the H$\alpha$ line (Karachentsev \& Makarov 1998), which
shows a rotation velocity of about 12 km/s at a radius of 20 arcseconds
(650 pc, at our distance),  
the formula 
$M_{\rm T} = V^2(R)RG^{-1}$
yeilds a lower limit of $3.4\times10^7 {\rm M}_\sun$ --- approximately 
the sum of the stellar and gas masses.  
More precise mass estimates, and in particlar a solid estimate of the
dark matter content, 
await new observations. 

Combining these estimates of the star formation history and total gas 
content of UGCA~290, we can also address its gas consumption timescale.
A very short gas consumption timescale implies that the  
current level of star formation cannot
be maintained for a long period, and is one of the
traditional arguments for the unusual nature of BCDs.  For UGCA~290,
the maximum burst level of $\sim 0.04~M_\sun~{\rm yr}^{-1}$ would take
600 million years to use up 
its HI gas mass, even at 100\% efficiency.  In fact, this ``burst" level was
maintained for only $\sim 15$ Myr.  The average star formation rate 
of UGCA~290 over 
the past billion years,  $\sim 0.003~M_\sun~{\rm yr}^{-1}$,  could be 
maintained for over seven billion years.  Therefore, UGCA~290 may well
continue forming stars in a similar manner for a long time. 

Thanks to the similarity between the bright stellar contents in UGCA~290 and
VII~Zw~403, we can compare our results for the recent burst
to observations of the latter galaxy as well.  
Recall that the bright stars in the V, I CMDs of these two galaxies 
(including Boxes A through D, and the main sequence down
to $M_{\rm I}=-4$), have essentially the same distribution.  The {\it total 
number} of stars in this region is nearly the same also, VII~Zw~403
having 96\% the number of stars in UGCA~290.  
Because the models were constrained using regions which are not
significantly different in the two galaxies, the results for UGCA~290
are consistent with the VII~Zw~403 data as well.  
(Compare the histograms in Figure 3 with those in Figure 7). 
There are two pieces of information which 
are {\it not} included in these statistical comparisons, 
which suggest that the current rate is slightly higher in VII~Zw~403: the
higher proportion of very hot stars revealed in F336W, and the higher
numbers of main sequence stars in the very lowest magnitude bins 
(which are, however, 
very
sensitive to an accurate determination of completeness).  
The current SFR for VII~Zw~403 may therefore be on the high side, 
but should still be within
the uncertainties we determined from our models. 
From the continuum-subtracted F656N images of VII~Zw~403 (part of 
the same observations described in Schulte-Ladbeck et al. 1999), 
we find a flux of $7.6\times 10^{-13}$ erg s$^{-1}$ cm$^{-2}$.  For 
a distance of 4.4 Mpc and 0.1 mag extinction, this corresponds to 
a luminosity $1.9\times 10^{39}$ erg s$^{-1}$, or a star formation
rate of 0.014~M$_\sun{\rm yr}^{-1}$.   This rate does agree
with our value of $0.011 \pm 0.008~{\rm M}_\sun{\rm yr}^{-1}$ 
from the resolved stars. 
Lynds et al. modeled the intermediate and old  
star formation history of VII~Zw~403 and concluded that the heavy 
AGB above the TRGB corresponds to a major episode of star formation which
lasted for $200-600$ Myr and ended about 600 Myr ago.  In the case of 
a 200 Myr burst, the burst level would be about 30 times the level
after the burst.  We emphasize that this burst does 
not have to be even this intense if
it is longer.  We also emphasize that the current burst --- the burst which
gives VII~Zw~403 its BCD appearance --- is a much shorter and more
recent phenomenon.

The similarity of the recent bursts in UGCA~290 and VII~Zw~403 
poses the question of whether they
have the same physical cause. 
Indeed, the causes of 
BCD bursts in general are not clear.  
Pustilnik et al. (2001) conclude
that most BCDs are triggered by tidal interactions or mergers, 
although VII~Zw~403 is one of the exceptions in their sample. 
Unfortunately, it is often difficult to judge whether a galaxy  
might be interacting
because of poor distance determinations,
difficulty in observing faint companions, and ambiguous merger morphologies. 
Even in the case of UGCA~290, where we can study the spatial distribution
of burst stars in detail, it is not clear what caused the burst. 
One might interpret its fairly moderate nature 
as support for the idea that we are 
simply catching it in
a relatively active phases of an internal stochastic process, but
this is not necessarily so. 
 
The behavior of these two galaxies 
fits neither of the scenarios outlined
in Searle and Sargent;  the recent bursts are not especially intense or brief,
and
most of their star formation occurred more than a billion years ago.
In addition, the
IMF slope appears to be close to Salpeter, at least for the massive stars.
Our growing knowledge of BCDs suggests that we should switch to a picture
of moderate enhancements instead of major bursts --- a picture 
that would solve the problem of the missing ``quiescent" counterparts, which need
not be so very different from BCDs themselves. 
Our results highlight, rather than solve, another part of the BCD puzzle: 
the cause of the recent enhancements in star formation. 
Are they triggered externally, or simply part of an internal stochastic
process?  What intrinsic properties are required of a gas-rich dwarf for
it to take part is such events? 

\acknowledgments We thank I. Karachentsev for providing us with 
HI data for UGCA~290.  Work on this project was supported through HST grants to RSL and 
MMC (project 8122), and by the Charles Lubin Chair at Skidmore College. 
We used NASA/IPAC Extragalactic Database, operated by the Jet Propulsion
Laboratory, California Institute of Technology, under contract
with NASA.

\clearpage

\begin{figure}
\caption{WFPC2 Planetary Camera image, showing the star-forming region of 
UGCA~290.  The inset,  our R-band image taken with the Calar-Alto 1.23 m
telescope, shows the WFPC2 field of view.  For both images, north is
to the lower right. 
} 
\end{figure}

\begin{figure}
\caption{Errors from DAOPHOT (top row), 
completeness fractions (middle row), and errors from false star
tests (bottom row) 
for our data in the F814W, F555w, and F336W filters.  For the
completeness fractions, 
solid lines show results for the PC chip and dashed lines show results
for the quarter of the WF2 chip nearest the main body of UGCA~290. 
For the error plots, the results for both chips are superimposed;
the two chips have effectively the same distribution of errors at a
given magnitude.} 
\end{figure}

\begin{figure}
%\epsscale{1.2}
%\vskip -1.0in
\caption[]{Color-magnitude diagrams in V and I for UGCA~290 and 
VII~Zw~403 (top) and their I-band luminosity functions. 
Color-coding in the CMDs indicates F336W-F555W color, as explained in
the text; stars not detected in F336W are left black.  Dashed 
lines indicate the regions we compare using the K-S test. 
The diagrams for both galaxies are set to the same absolute magnitude scale.
The apparent magnitude scale on the left refers to 
UGCA~290; the distance modulus for VII~Zw~403 is 0.9 mag less.
The lower panels show the luminosity functions for both 
UGCA~290 (bold lines) and 
VII~Zw~403 (fine lines), corrected for incompleteness.  
On the left are 
$\sqrt N$ error bars for reference. } 
\end{figure}
%\clearpage

\begin{figure}
%\epsscale{0.8}
%\plotone{f4.eps}
%\vskip 1.0in
\caption{Color-magnitude diagrams in F336W and F555W (top), 
and color-color diagrams (bottom). 
The color-coding indicates I-band magnitude, as explained in the text. 
All the stars detected in both F336W and F555W were also detected
in $I$.  The dotted line in the upper panels show the regions we compare
using the K-S test.
The arrows in the lower panels are reddening vectors for E($B-V$)=0.5 mag.
} 
\end{figure}
%\clearpage

\begin{figure}
%\vskip 1.0in
%\plotone{f5.eps}
%\vskip -1.0in
\caption{The PC data superposed with several Padova evolutionary
tracks (left) and with boxes used for creating synthetic models (right). 
The tracks include models with Z=0.004 (red) and Z=0.0004 (blue) for
masses 30 M$_\sun$, 12 M$_\sun$, and
1 M$_\sun$.} 
\end{figure}

\begin{figure}
%\vskip 1.0in
%\epsscale{1.2}
%\plotone{f6.eps}
%\vskip -2.0in
\caption{The recent star formation history of UGCA~290,
according to our best model.} 
\end{figure}

\begin{figure}
%\epsscale{1.3}
%\plotone{f7.eps}
%\vskip -1.0in
\caption{The best model, along with the PC data
for comparison.  Below are the luminosity functions for the 
data (bold lines) and the model (fine lines). 
On the left are 
$\sqrt N$ error bars for reference. }
\end{figure}

\begin{figure}
%\plotone{f8.eps}
%\vskip -1.0in
\caption{The best low-metallicity model, as described in the text.
The format is the same as that for Figure 7.} 
\end{figure}

\begin{figure}
%\plotone{f9.eps}
%\vskip -1.0in
\caption{The best model with a steep IMF exponent $\alpha=3.00$}
\end{figure}

\begin{figure}
%\plotone{f10.eps}
%\vskip -1.0in
\caption{The best model with a flat IMF exponent $\alpha=1.35$} 
\end{figure}

\begin{figure}
%\plotone{f11.eps} 
%\vskip -1.0in
\caption{Same as the model in Figure 7, 
except with the recent burst beginning 50 Myr ago
instead of 20 Myr ago.} 
\end{figure}

\begin{figure}
%\plotone{f12.eps} 
%\vskip -1.0in
\caption{Same as the model in Figure 7, 
except without a recent burst.  Instead,
star formation is constant from 50 Myr ago onward.} 
\end{figure}

\begin{figure}
%\plotone{f13.eps}
%\vskip -1.0in
\caption{Same as the model in Figure 7, 
except with an additional burst  
at 400 Myr with duration 20 Myr and rate 0.04 M$_\sun {\rm yr}^{-1}$.  
The loss of time resolution this long ago is such that the burst 
is nearly hidden.} 
\end{figure}

\begin{figure}
%\plotone{f14.eps}
%\vskip -1.0in
\caption{Same as the model in Figure 7, except with a gap in
star formation from $500-1000$ Myr ago.}  
\end{figure}

\begin{figure}
%\vskip 1.0in
%\plotone{f15.eps}
%\vskip -4.0in
\caption{Data from the WF chips.  The
quarter of the WF2 chip closest to the PC (a) includes primarily stars in the
red background sheet of UGCA~290 and is considered part of that galaxy.
The quarter of the WF3 chip farthest from the PC (b) gives an estimate of 
contamination by objects not in UGCA~290.} 
\end{figure}

\begin{figure}
%\epsscale{0.9}
%\plotone{f16.eps}
\caption{Synthetic models of the WF data.  On the left are
color-magnitude diagrams.  On the right are the distributions in
color in the region $0.8<(V-I)_o<2.3$, $-4<M_I<-3$ 
(the tip of the red giant branch.) See text for the model parameters.} 
\end{figure}

\begin{figure}
%\epsscale{0.95}
%\plotone{f17.eps}
%\vskip -2.0in
\caption{Distribution of stars with progressively older ages. 
From left to right, the age ranges are approximately $0-20$ Myr, 
$20-50$ Myr, $50-1000$ Myr, and $> 1000$ Myr.  Solid lines indicate
the WFPC2 field of view.} 
\end{figure}

\begin{figure}
%\vskip 0.5in
%\epsscale{0.95}
%\plotone{f18.eps}
%\vskip -1.0in
\caption{Distribution of stars in the recent burst, along with
H$\alpha$ contours. The vertical line on the left shows a scale of 
1 kpc.} 
\end{figure}
\clearpage

\begin{figure}
%\plotone{f19.eps} 
%\vskip -2.0in
\caption{The V,I CMD for each of the two major spatial lobes of 
star formation. } 
\end{figure} 

\end{document}